\documentclass[journal]{IEEEtran}
\usepackage{amsthm}
\usepackage{tabularx}
\usepackage{cite}
\usepackage[dvips]{graphicx}
\usepackage[table, dvipsnames]{xcolor}
\usepackage{color, colortbl}
\definecolor{lavenderblue}{rgb}{0.8, 0.8, 1.0}
\definecolor{skyblue}{rgb}{0.53, 0.81, 0.92}
\definecolor{blizzardblue}{rgb}{0.67, 0.9, 0.93}
	\definecolor{aqua}{rgb}{0.0, 1.0, 1.0}
   \graphicspath{{../eps/}}
   \DeclareGraphicsExtensions{.eps}

\usepackage{mathrsfs}
\usepackage{amssymb}
\usepackage{yfonts}
\usepackage[cmex10]{amsmath}
\usepackage{algorithmicx}
\usepackage{algpseudocode}
\usepackage{algorithm}
\usepackage{graphicx}
\usepackage{array}
\usepackage{color}
\usepackage[caption=false,font=footnotesize]{subfig}
\usepackage{lettrine}
\usepackage{tikz}
\usepackage{verbatim}

{}
\floatname{algorithm}{\textbf{Algorithm}}{}
\ifCLASSINFOpdf

\else

\fi
\begin{document}
\title{A lightweight Encryption Method For Privacy-Preserving in Process Mining}

\author{Mohsen~Kazemian,~\IEEEmembership{Member,~IEEE,}
        Markus~Helfert
\thanks{M. Kazemian is with the Adapt Research Centre and Innovation Value Institute (IVI), Maynooth University, Maynooth, Ireland (e-mail: mohsenkazemian@yazd.ac.ir).}

\thanks{M. Helfert is with the Adapt Research Centre, Innovation Value Institute and School of Business, Maynooth University, Ireland
(e-mail: markus.helfert@mu.ie).}

}

\maketitle

\begin{abstract}
  Novel technological achievements in the fields of business intelligence, business management and data science are based on real-time and complex virtual networks. Sharing data between a large number of organizations that leads to a system with high computational complexity is one of the considerable characteristics of the current business networks.
  Discovery, conformance and enhancement of the business processes are performed using the generated event logs. In this regard, one of the overlooked challenges is privacy-preserving in the field of process mining in the industry. To preserve the data-privacy with a low computational complexity structure that is a necessity for the current digital business technology, a novel lightweight encryption method based on Haar transform and a private key is proposed in this paper. We compare the proposed method with the well-known homomorphic cryptosystem  and Walsh-Hadamard encryption (WHE) in terms of cryptography, computational complexity and structure vulnerability. The analyses show that the proposed method anonymizes the event logs with the lower complexity and more accuracy compared with two aforementioned cryptosystems, significantly.
\end{abstract}


\IEEEpeerreviewmaketitle

\section{Introduction}
Advent of $5$G and beyond technologies provides many fascinating possibilities for all the companies and organizations to make their smart and digital business come to reality \cite{bus}. These advancements aim to transfer the demanded information based on the digital environments. Carrying the information between machines and manufacturing execution system, and between warehouse robots and related management system in the industry, are two examples in this domain \cite{ieeerobot}. Therefore, mining the communications in any organisation with the aim of improving the executed processes is vital in the current digital world.

Process mining (PM) refers to a family of methods focused on getting valuable insights from executed processes in any organization using the generated information \cite{spm}, \cite{securitypm}. Process mining techniques are related to the fields of data science (i.e. areas such as data mining and predictive analytic) and process science (e.g. business process management (BPM) \cite{bpmn}, \cite{prof}) having the aim of discovering the bottlenecks and improving the overall system performance based on the event logs.

Currently, process mining techniques have been widely developed in industry and academia. In this regard one of the most important domains is healthcare enabling healthcare stakeholders to get valuable insights such as identifying the actual order of activities and the involvement of resources, from an event log \cite{healthcarezerotrust}. An event log in healthcare may include the sensitive and critical information of the patient such as the case ID, activity, time stamps, age, diagnosis, and treatment code.

The need to consider privacy-preserving in process mining was felt at an early stage in the studies \cite{manifesto}. However, the process mining association has typically overlooked the privacy problem, until recently. Currently, with the emergence of real-time and virtual network architectures, more than before, adequately safeguarding data privacy and security is of key importance for process mining area. In this domain, one of the essential targets is to encrypt the data log so that in the presence of untrusted environments, interfaces and organizations, the information must be kept secure before applying the process mining algorithms \cite{towardpm}. Since the current emerging networks are much complex inherently, development of a privacy-preserving approach using a low complexity algorithm is of key importance. Given the point that considering the variable parameters such as security degree, hardware architecture, processing speed and the volume of data are key parameters in the selection of encryption method and process mining technique.
\subsection{Related Work}
Privacy and security-preserving topics are entirely well-known in the general field of data mining. Currently, some articles made preliminary efforts to address some noticeable challenges related to privacy-preserving in process mining science \cite{manifesto}, \cite{GG}. A method allowing the outsourcing of process mining while ensuring the secrecy of event logs and the executed processes is proposed in \cite{towardpm}. This work offers a very weak privacy-protection approach and could be prone to advanced de-anonymization schemes. Saavedra et al. \cite{mask} proposed a model aims to mask event logs containing sensitive values (e.g. by removing the last $4$ characters from $\{12345678\}$). However, some drawbacks such as high operational complexity and the uncertainty in regard to the correct regeneration of the original data before applying the process mining techniques, make it unacceptable. Liu et al. \cite{liu} proposed a privacy-preserving framework for cross-organization business based on access control. Process mining with encrypted data is one of the main targets in this research area. However, to the best of our knowledge, this goal has not yet been achieved. In this regard homomorphic encryption (HE) \cite{homo} designed to allow arithmetic operations to be performed on the encrypted data. Therefore, designing a process mining algorithm that works with homomorphic encrypted data is a novel research area. However, HE with its very high computational complexity and high delay architecture is not a general solution for all applications. Furthermore, the general process mining algorithms are not designed to work with HE, efficiently. Hence, existence of a low computational complexity approach that works with the current PM algorithms is admirable.
\subsection{Main Contributions}
In this paper we propose a novel approach called Haar transform encryption (HTE) with a low computational complexity structure. HTE is designed for both numbers and characters using the simple architecture of the Haar matrices \cite{HAAR} and one private-key. Therefore, for the networks and processors considering multiplication as a time-consuming operation, a significant saving is achieved. Since process mining techniques are independent of domain and can be applied in any organization where processes are present and a data log is available, the proposed approach is applicable to any industry. However, it is analyzed on the healthcare domain through the paper. Addressing the important necessities of the current digital business networks such as low computational complexity and easy-restorable structure of the cryptographic technique for the event-log data in a novel designed system model are the aims of this study.

Furthermore, the computational complexity of the Paillier technique as one of the well-known homomorphic cryptosystems is studied to prove that the improvement of this cryptosystem is an essential requirement if there is a process mining algorithm that works with the encrypted data. Note that our proposed symmetric cryptographic technique needs to be decrypted before applying the process mining algorithms that is suitable for the current applications.

The rest of the paper is organized as follows: Section \ref{defin1} presents some preliminary definitions including the Homomorphic and Haar transform concepts. The system model and the proposed scheme are introduced in Section \ref{defin1.5} and \ref{defin2}, respectively. Results and discussion including encryption of an event log and the computational complexity analysis are presented in Section \ref{defin3}, and is followed by the conclusion in Section \ref{defin4}.

\textit{Notation:} Throughout the paper, we use small and capital boldface letters, $\otimes$, $\alpha^{T}$ and $\{\cdot\}$ to denote vectors, matrices, Kronecker product, the transpose of $\alpha$ and multiplication operator, respectively.
\section{Definitions} \label{defin1}
The preliminary definitions through the proposed method are defined as follows:

\emph{Definition 1}~~(Haar transform):
Un-normalized Haar unitary matrices consist of $\pm1$ and $0$, and the transforms are provided only by addition and subtraction operations, without involving any multiplications. However, the Haar transform uses the normalized Haar matrices including the other numbers in addition to $\pm1$ and $0$ \cite{HAAR}. Although multiplication operations are needed in the Haar transform, the computation time is very short because of the existence of $\pm1$ and $0$ values in its matrix structure. Therefore, a significant saving in terms of complexity and delay is achieved.
Haar functions $\hbar_{\upsilon}(\varphi)$ are defined in the interval $[0, 1]$ and the order $\upsilon$ of the function is uniquely decomposed into the integers $a$ and $b$ as follows:
\begin{equation}
\upsilon = {2^a}+b-1,~~ N=2^l,~~\upsilon\in\{0,1,...,N-1\},
\end{equation}
where $ l \in \{1,2,...\}$, $0\leq a \leq l-1$, $0\leq b\leq{2^a}$. The Haar functions are defined as follows:
\begin{equation}
\hbar_{0}(\varphi) \equiv \hbar_{00}(\varphi)= \frac{1}{\sqrt{N}},~~\varphi\in [0, 1],
\end{equation}
\begin{equation}
\hbar_{\upsilon}(\varphi) \equiv \hbar_{ab}(\varphi)= \frac{1}{\sqrt{N}}
\begin{cases}
  ~2^{\frac{a}{2}}  & \frac{b-1}{2^a}\leq \varphi < \frac{b-{\frac{1}{2}}}{2^a}, \\  -2^{\frac{a}{2}} & \frac{b-\frac{1}{2}}{2^a}\leq \varphi < \frac{b}{2^a}, \\  ~0 & \mbox{otherwise}\mbox{ in [0,1]}.
\end{cases}
\end{equation}

The Haar transform matrix of order $N$ consists of rows which are resulted from the prior functions computed at the points $\varphi=\frac{e}{N}$ where $e\in\{0,1,...,N-1\}$. Considering $\mathbf{H}$ as Haar matrix which is a square matrix of dimension $2^l$, the vector $\mathbf{\check{y}}$ as the Haar transform of an $N$-point vector $\mathbf{x}$ is computed by:
\begin{equation}\label{equ4}
\mathbf{\check{y}}= \mathbf{H}_{2^{l}}\mathbf{x}.
\end{equation}

Note that the Haar matrix generally can be derived by the following equations:
\begin{equation}\label{eq:5}
\mathbf{H}_{2^{l}}=\begin{bmatrix}
\mathbf{H}_{{2^{l-1}}}\otimes [1, 1]\\
\mathbf{I}_{{2^{l-1}}}\otimes [1, -1]\\
\end{bmatrix},
\quad
\end{equation}
\begin{equation}\label{eq:6}
\mathbf{I}_{{2^{l-1}}}=\begin{bmatrix}
1 & 0 & \mathbf{\hdots} & 0\\
0 & 1 & \mathbf{\hdots} & 0\\
\mathbf{\vdots} & \mathbf{\vdots} & \mathbf{\ddots} & \mathbf{\vdots}\\
0 & 0 & \mathbf{\hdots} & 1\\
\end{bmatrix}_{{{2^{l-1}}\times {2^{l-1}}}}.
\quad
\end{equation}


\emph{Definition 2}~~(Walsh-Hadamard (WH)):
The square matrix of WH with dimension $2^l$ is defined as follows \cite{KAZEMIAN}:
\begin{equation}\label{eq:70}
\mathbf{\bar{H}}_{2^{l+1}}=\mathbf{\bar{H}}_{2^{l}}\otimes \mathbf{\bar{H}}_{2}=\begin{bmatrix}
\mathbf{\bar{H}}_{2^{l}} & \mathbf{\bar{H}}_{2^{l}}\\
\mathbf{\bar{H}}_{2^{l}} & -\mathbf{\bar{H}}_{2^{l}}\\
\end{bmatrix},
\quad
\end{equation}
with
\begin{equation}
\mathbf{{\bar{H}}}_{2}=\begin{bmatrix}
1 & 1\\
1 & -1\\
\end{bmatrix}.
\quad
\end{equation}

This matrix can also be used for the encryption strategy. However, absence of zero element in the Walsh-Hadamard matrix makes it more complicated than the Haar one.

\emph{Definition 3}~~(Homomorphic encryption):
HE allows arithmetic operations including addition and multiplication over encrypted data without decryption procedure which can be used as a basis for computing complex functions \cite{homo}. Two types of HE have received more attention: partial homomorphic and fully homomorphic encryption cryptosystems. The former allows only one operation either addition or multiplication, while the latter supports both addition and multiplication operations to be performed on the ciphertexts (CTs) with the aim of obtaining the computational results on the corresponding plaintexts (PTs).

One of the most recent invented cryptosystems is paillier homomorphic encryption (PHE) which is only an additive cryptosystem \cite{paillier}. In the following, the workflow of paillier system is explained.

The paillier algorithm is started by randomly choosing two independent large prime numbers $j$ and $k$ when $gcd~(jk, (j- 1)(k-1)) = 1$. Computing $n=jk$, $\lambda = lcm~(j-1, k-1)$ and selecting random integer $g$ where $g$ $\in$ $\mathbb{Z}_{n^2}^*$ are the next steps.
The above steps must be repeated until it is confirmed that $n$ divides the order of $g$. This is done by checking the existence of $\nu = {(L ({g^\lambda}  \mod  {n^2}))^{-1}} \mod n$ where $L(u) = \frac{u-1}{n}$. In this situation, the pair $(n,g)$ and $(\lambda,\nu)$ are known as public and private keys, respectively.

If $x \in \mathbb{Z}_{n}$ and $r \in \mathbb{Z}_{n}^*$ are the plaintext message and a random number, respectively, then $E(x)={g^x}.{r^n}\mod {n^2}$ is the ciphertext,  where $E(x) \in \mathbb{Z}_{n^2}^*$. The plaintext will be decrypted using the following equation: $x = {L ({{E(x)}^\lambda}  \mod  {n^2})\cdot {\nu}} \mod n$. Since paillier cryptosystem is just an additively homomorphic cryptosystem, the product of two ciphertexts is decrypted to the sum of their corresponding plaintexts as follows:
\begin{equation}
D(E({x_1},{r_1})\cdot E({x_2},{r_2}) \mod {n^2}) = {x_1} + {x_2} \mod n,
\end{equation}
where $D(E({x}))=x$. If $g$, $r$, $n$ and specially the plain text $x$ are large values, then a huge computational complexity and delay will be occurred in both hardware and software. Hence, HE with the aforementioned structure including large number of multiplication operations, is not suitable for current high speed frameworks. A summary of PHE is described in Algorithm \ref{algo1}.
\begin{algorithm}[]
\caption{Paillier Homomorphic Encryption Procedure}
\textbf{Input:} $\mathbf{x}=[x_{1},...,x_{N}]^T$ for $N$ numbers.
\begin{algorithmic}[1]
\For{$i=1,...,N$}
\State {\textbf{repeat}}
\State~~~~{generate $j$ and $k$ when $gcd~(jk, (j-1)(k-1)) = 1,$}
\State ~~~~{set $n       \leftarrow jk,$}
\State ~~~~{set $\lambda \leftarrow lcm(j-1,k-1),$}
\State~~~~{select random $g$ where $g$ $\in$ $\mathbb{Z}_{n^2}^*$,}
\State {\textbf{until}} {${gcd~(L ({g^\lambda}  \mod  {n^2}),n)}\neq1$,}

\State{select random $r_i$ where $r_i \in \mathbb{Z}_{n}^*$,}
\State{compute ciphertext $E(x_i) = {g^{x_i}} . {{r_i}^{n_i}} \mod {{n_i}^2},$}
\EndFor
\end{algorithmic}
\textbf{Output:}   $\mathbf{E(x)}=[E{x_{1}},...,E{x_{N}}]$, $\mathbf{r}=[{r_{1}},...,{r_{N}}]$.
\label{algo1}
\end{algorithm}
\section{System Model}\label{defin1.5}
\begin{figure} [t]
\centering \includegraphics [scale = 0.74]{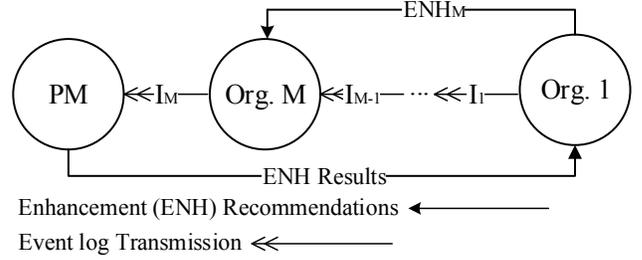}
\caption{System model for $M$ organizations}\label{fig:1}
\end{figure}

Beyond Industry ${4.0}$ exploits the collaboration of multi organizations in serial and parallel situations even at different geographical points, where the data transfer with untrusted environment is a serious concern \cite{infoletter}. In this regard, we have considered a digital business network consisting
of {$M\in\mathbb{N}$} organizations and one process mining point, as shown in Fig. \ref{fig:1}. The event log is created at the {$1^{st}$} organization (Org. $1$) and will be updated at the subsequent $M-1$ organizations by adding new data. PM point applies the process mining algorithms and sends the results to the first organization. Finally, the Org. {$1$} sends the enhancement instructions to the subsequent organizations. In this scenario, Org. {$1$} and the PM point are trusted, and the ${I_m}$, ${{m}\in\{1,...,M\}}$, which is any type of interface to transmit the event log to the next organization is the untrusted environment. Note that any organization, except the first one, is considered as untrusted environment for the other organizations' data. In other words, the data of each organization is encrypted using an individual private key.
\section{Proposed Method}\label{defin2}
Let $\mathbf{y}=[y_{1},...,y_{N}]^T$ denote the cipher vector of $\mathbf{x}=[x_{1},...,x_{N}]^T$, then using (\ref{equ4}) we have:
\begin{equation} \label{hte}
\begin{bmatrix}
y_{1}\\
y_{2} \\
\vdots \\
y_{N,} \\
\end{bmatrix}
\quad= {\zeta_p}\frac{1}{\sqrt{N}}{{\mathbf{H}_{2^{l}}}}\begin{bmatrix}
x_{1}\\
x_{2} \\
\vdots \\
x_{N} \\
\end{bmatrix},
\quad
\end{equation}
where ${\zeta_p}\in \mathbb{R}$ is the arbitrary random private key which is considered in the encryption and decryption procedure, where ${{p}\in\{m,s\}}$. Individual private key ${\zeta_m}$ makes the data of {$m^{th}$} organization inaccessible from the untrusted environments, when ${\zeta_s}$ refers to the shared key between all organizations.

Since $\mathbf{H}_{2^{l}}$ is a square matrix of order $2^l$ \cite{HAAR}, if the actual number of data denoted by $\bar{N}$ is less than ${2^l}$, then $\tilde{N}={2^l}-\bar{N}$ zeros are padded to the data vector. For instance, for a small data log with five numbers, we need to pad three zeros to adapt with an $8\times8$ Haar matrix. Finally, using the fact that for the Haar matrix $\mathbf{H}^{-1}_{2^{l}}=\mathbf{H}_{2^{l}}^{T}$, the vector $[x_{1},...,x_{N}]^T$ can be extracted from $\mathbf{y}$ as follows:
\begin{equation} \label{eqb10}
\begin{bmatrix}
x_{1}\\
x_{2} \\
\vdots \\
x_{N} \\
\end{bmatrix}
\quad=         \frac{1}{\zeta_m} \frac{1}{\sqrt{N}}{{\mathbf{H}_{2^{l}}^{T}}}\begin{bmatrix}
y_{1}\\
y_{2} \\
\vdots \\
y_{N} \\
\end{bmatrix}.
\quad
\end{equation}
\begin{table*}[t]
  \begin{center}
    \caption{Encryption of an event-log in healthcare domain using the proposed method in Org.$1$ and Org.$2$}
    \label{tab:1}
    \begin{tabular}{cc|cc|cc|cc|cc} 

      \textbf{Case ID} && \textbf{Timestamp} && \textbf{Activity} && \textbf{Resource} && \textbf{Heart rate} \\
      \hline

      \textbf{PT} & \textbf{CT ($\zeta_s$)} & \textbf{PT}& \textbf{CT ($\zeta_s$)} & \textbf{PT} & \textbf{CT ($\zeta_s$)}& \textbf{PT} & \textbf{CT ($\zeta_1$)} & \textbf{PT} & \textbf{CT ($\zeta_2$)}\\
      \hline
      1 & $12$ & $10:20$ & $6812$& {$A\rightarrow 1$}& $23$ & $Tom$ & ($48,22,7,18.2$) & $72$ & $293$ \\
  2 & $0$ & $11:31$ & $-1286$& {$A\rightarrow 1$}& $-11$ & $Tom$ & ($48,22,7,18.2$) &  $78$ & $-8$ \\
  1 & $0$ & $11:42$ & $-199.40$& {$B\rightarrow 2$}& $-2.82$ & $Tom$ & ($48,22,7,18.2$) &  $60$ & $10.6$ \\
  2 & $-2.82$ & $12:30$ & $-535.98$& {$B\rightarrow 2$}& $-1.41$ & $Tom$ & ($48,22,7,18.2$) & $75$ & $-45.96$ \\
  1 & $-2$ & $14:10$ & $-142$& {$C\rightarrow 3$}& $0$ & $John$ & ($47,2,-7,-8.4$) &  $58$ & $-6$ \\
  1 & $-2$ & $16:25$ & $-96$& {$E\rightarrow 5$}& $0$ & $John$ & ($47,2,-7,-8.4$) & $60$ & $-15$ \\
  $2$ & $0$ & $17:30$ & $-270$& {$D\rightarrow 4$}& $-4$ & $Anna$ & ($30,0,-18.2,18.2$) & $90$ & $-2$ \\
  $2$ & $0$ & $19:24$ & $-228$& {$E\rightarrow 5$}& $-2$ &$Anna$ & ($30,0,-18.2,18.2$) &  $93$ & $-3$ \\

    \end{tabular}
  \end{center}
\end{table*}

\begin{table*}[t]
  \begin{center}
    \caption{Available part of the event log to Org.$3$}
    \label{tab:2}
    \begin{tabular}{cc|cc|cc|cc|cc} 

      \textbf{Case ID} && \textbf{Timestamp} && \textbf{Activity} && \textbf{Resource} && \textbf{Cost {(\$)}} \\
      \hline

      \textbf{PT} & \textbf{CT ($\zeta_s$)} & \textbf{PT}& \textbf{CT ($\zeta_s$)} & \textbf{PT} & \textbf{CT ($\zeta_s$)}& \textbf{PT} & \textbf{CT ($\zeta_1$)} & \textbf{PT} & \textbf{CT ($\zeta_3$)}\\
      \hline
      1 & $12$ & $10:20$ & $6812$& {$A\rightarrow 1$}& $23$ & $Tom$ & ($48,22,7,18.2$) & $3$ & $126$ \\
  2 & $0$ & $11:31$ & $-1286$& {$A\rightarrow 1$}& $-11$ & $Tom$ & ($48,22,7,18.2$) &  $3$ & $-54$ \\
  1 & $0$ & $11:42$ & $-199.40$& {$B\rightarrow 2$}& $-2.82$ & $Tom$ & ($48,22,7,18.2$) &  $3$ & $0$ \\
  2 & $-2.82$ & $12:30$ & $-535.98$& {$B\rightarrow 2$}& $-1.41$ & $Tom$ & ($48,22,7,18.2$) & $3$ & $-42$ \\
  1 & $-2$ & $14:10$ & $-142$& {$C\rightarrow 3$}& $0$ & $John$ & ($47,2,-7,-8.4$) &  $5$ & $0$ \\
  1 & $-2$ & $16:25$ & $-96$& {$E\rightarrow 5$}& $0$ & $John$ & ($47,2,-7,-8.4$) & $5$ & $0$ \\
  $2$ & $0$ & $17:30$ & $-270$& {$D\rightarrow 4$}& $-4$ & $Anna$ & ($30,0,-18.2,18.2$) & $10$ & $0$ \\
  $2$ & $0$ & $19:24$ & $-228$& {$E\rightarrow 5$}& $-2$ &$Anna$ & ($30,0,-18.2,18.2$) &  $10$ & $0$ \\

    \end{tabular}
  \end{center}
\end{table*}
Any event log consists of data attributes in three types of numbers, timestamps and characters. The last two types have to be mapped into the unique numbers as described below:

\textbf{Timestamps} are calculated from a time-origin and the differences in minutes or seconds are the numerical expressions.

\textbf{Characters} are mapped into numbers based on their alphabetical order.

We summarize the design steps of the proposed scheme as the pseudocode in Algorithm \ref{algo2}. To get more insight into how this algorithm works, lines $1$ to $3$ of Algorithm \ref{algo2} deal with the mapping the characters and timestamps to the integer numbers. If the actual number of data vector members, denoted by $\bar{N}$, is less than $2^l$, then the zero padding procedure is required. (see lines $4$ to $6$ of Algorithm \ref{algo2}). The encrypted vector $\mathbf{y}$ is simply calculated by line $8$. Finally, $\mathbf{y}$ and $\zeta_p$ are used for the decryption procedure before applying the process mining tools.
\begin{algorithm}[t]
\caption{Proposed HTE Scheme for the $m^{th}$ Organization}
\textbf{Input:} $\mathbf{x}=[x_{1},...,x_{N}]^T$ for $\bar{N}$ numbers.
\begin{algorithmic}[1]

\If{$\mathbf{x}_i \notin \mathbb{R}$} \State{map to numbers}
\EndIf

\If{$\bar{N}<2^l$} \State{perform $\tilde{N}=2^l-\bar{N}$ zero padding procedure,}
\State{set $N \leftarrow \tilde{N}+\bar{N}$.}
\EndIf
\State{calculate $\mathbf{y}=\zeta_m\frac{1}{\sqrt{N}}\mathbf{H}_{2^{l}}[x_{1},...,x_{N}]^T$,}

\end{algorithmic}
\textbf{Output:} $\mathbf{y}$, $\zeta_m$.
\label{algo2}
\end{algorithm}
\begin{table*}[t]
  \begin{center}
    \caption{Complexity comparison between HTE, WHE and PHE based on two data vectors from Table~\ref{tab:1} in terms of required multiplication and addition numbers}
    \label{tab:3}
    \begin{tabular}{cc|cc|cc|cc} 

      \textbf{Data Vector} &&\textbf{HTE} &&\textbf{WHE} && \textbf{PHE} \\
      \hline

       $\textbf{x}$ && \textbf{Mul.}& \textbf{Add.} & \textbf{Mul.}& \textbf{Add.} & \textbf{Mul.}& \textbf{Add.}  \\
      \hline
      Heart rate && $32$ & $24$ & $64$ & $56$ & $(22^{72}+22^{78}+22^{60}+22^{75}+22^{58}+22^{60}+22^{90}+22^{93})-8$ & $0$ \\
  Case ID && $32$ & $24$ & $64$ & $56$ & $(22^{1}+22^{2}+22^{1}+22^{2}+22^{1}+22^{1}+22^{2}+22^{2})-8$ & $0$ \\

    \end{tabular}
  \end{center}
\end{table*}

\section{Results and Discussion}\label{defin3}
In this section we investigate the manner of our proposed method on an event log with typical healthcare data attributes using $M=3$ organizations. Additionally, comparisons based on computational complexity and the vulnerability of the cryptographic structure for HTE, WH encryption (WHE) \cite{KAZEMIAN} and PHE schemes are studied. Throughout this section $\zeta_s$, $\zeta_1$, $\zeta_2$, $\zeta_3$, $j$, $k$ and $g$ are considered $\sqrt{8}$, $2$, $\sqrt{2}$, $3\sqrt{8}$, $3$, $5$ and $22$, respectively.
\subsection{Encryption using HTE}
A small event log together with its anonymized versions which are encrypted by HTE approach are described in Table~\ref{tab:1} and Table~\ref{tab:2}. The log is made of eight events, two distinct case IDs: $\{1, 2\}$, five different activities: $\{A, B, C, D, E\}$ which are performed by three users: \{Tom, John, Anna\}, in ascending time stamps. Org. $1$ uses $\zeta_s$ to encrypt the case ID, timestamp and activity columns, as the shared data between all organizations. Additionally, it uses $\zeta_1$ to anonymize the resource column to protect the privacy-preserving of the patients. Updating process can add data to both rows and columns of the event log. In this example, Org. $2$ only adds the encrypted column of heart rate information using $\zeta_2$, when it has no knowledge of the resource column. In a specific case, if needed, $\zeta_1$ can be available to Org. $3$ and the encrypted column of cost will be added to the event log using $\zeta_3$. Hence, any organization is able to use the data of the other organizations, provided it has the corresponding key. Table~\ref{tab:2} describes the available information for Org. $3$. Finally, the trusted PM point applies the process mining algorithms using the decrypted event log and send the enhancement recommendations to the $1^{st}$ organization. Collaboration of thousands organizations with large event logs is conceivable in the future digital business networks.

Referring to \ref{hte} and Algorithm \ref{algo2}, the Haar transform of numerical data vector $\mathbf{x}$ multiply by $\zeta_p$, is the encrypted vector $\mathbf{y}$. The time stamps are calculated from a time-origin and generate the numerical vector $\mathbf{x}$ (e.g. ${10 : 20}$ - ${00:00}= 620$ minutes). Note that each column in Table~\ref{tab:1} and Table~\ref{tab:2} has its plaintext vector $\mathbf{x}$ and corresponding ciphertext vector $\mathbf{y}$, exclusively.
In the non-numerical data attributes such as activity and resource, $\mathbf{x}$ is formed by mapping each character into a number based on its alphabetical order. However, if the member in the plaintext vector includes more than one character, then a separate $\mathbf{y}$ will be presented for each word (e.g. Tom $\rightarrow \mathbf{x}=[20, 15, 13, 0]$ and $\mathbf{y}= [48,22,7,18.2]$ with $N=4$). Finally, the ciphertext columns in Table~\ref{tab:1} and Table~\ref{tab:2} will be considered as the encrypted data log.
\subsection{Computational Complexity Analysis}\label{defin4}
In this subsection, we present an investigation on the computational complexity of the paillier cryptosystem and the proposed scheme, to show the superiority of the latter based on the number of multiplication operations. We prove that the overall order of PHE scheme described in
Algorithm \ref{algo1} is $\mathcal{O}\left({}{\sum_{i=1}^N g^{x_i}}\right)$ while that of HTE described in Algorithm \ref{algo2} is $\mathcal{O}\left({N^2}\right)$.

\emph{Proof 1:} There are three bottleneck lines in the complexity analysis of Algorithm \ref{algo1}, i.e., lines $1$ - $10$, $2$ - $7$ and $9$. The first bottleneck that includes a \textit{``for''} loop with the length of $N$, is of order $\mathcal{O}(N)$. Using the simplest assumption that makes the condition in line $7$ true at the first loop, the computational complexity of the second bottleneck including a \textit{``repeat''} loop is limited to $gcd$ and $lcm$ algorithms, which are of order $\mathcal{O}\left(\log{}{jk.{((j-1)(k-1))}}\right)\equiv\mathcal{O}\left(\log{}{n^2}\right)$ and $\mathcal{O}(1)$, respectively, where $lcm=\frac{(|{(j-1).(k-1)}|)}{gcd{(j-1,k-1)}}$. Considering $N\ll {x_i}$ and $n\ll {x_i}$ where $i\in\{1,...,N\}$, only the third bottleneck including power operation with the order of $\mathcal{O}\left({g^{x_i}}\right)$ forms the dominant term. Thus, the overall order of PHE in the most simple situation is $\mathcal{O}\left({}{\sum_{i=1}^N g^{x_i}}\right)$. Indeed, the complexity of PHE grows with dependency on the data value that yields huge computational complexity in the hardware and programming architecture.

\emph{Proof 2:} Referring to Algorithm \ref{algo2}, the only bottleneck is in line $8$ that includes the main part of the proposed scheme. Referring to (\ref{eq:5}) and (\ref{eq:6}), we extract ${z}({\mathbf{H}_{2^{l}}})$ as the number of zeros in ${\mathbf{H}_{2^{l}}}$ using the following equation:
\begin{equation}
{z}({\mathbf{H}_{2^{l}}}) = 2({2^{2(l-1)}}-{2^{l-1}})+2{z}({\mathbf{H}_{2^{l-1}}}),~~~ {z}({\mathbf{H}_{2}}) = 0.
\end{equation}

The actual number of multiplication and addition operations for a data vector with the length of $N=2^l$ using HTE scheme are achieved as follows:
\begin{equation}
Mul_{N} = {N^2}-{z}({\mathbf{H}_{2^{l}}}),
\end{equation}
\begin{equation}
Add_{N} = {{N^2}-N}-{z}({\mathbf{H}_{2^{l}}}).
\end{equation}

Therefore, a significant reduction in terms of multiplication numbers is achieved especially for the large ${x_i}$ and $N$ values, due to the exitance of zeros in Haar matrices. Finally, the overall order of the HTE scheme is computed by $\mathcal{O}\left({N^2}\right)$.

\begin{figure} [t]
\centering \includegraphics [scale = 0.63]{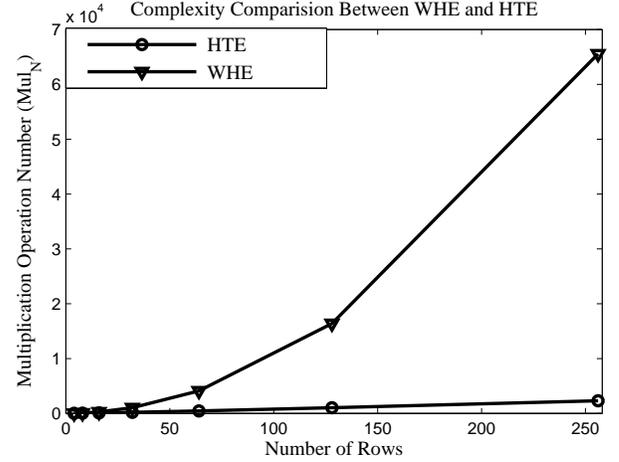}
\caption{Computational complexity comparison between WHE and HTE in terms of multiplication numbers for $N=4$ to $N=256$}\label{fig:2}
\end{figure}

A comparison based on the numbers of multiplication and addition operations between HTE, WHE and PHE based on the data in Table~\ref{tab:1}, is described in Table~\ref{tab:3}. Since the computational complexity of PHE is dependent on the values of $N$ and ${x_i}$, a large number of multiplication operations is required for the large data values while the proposed scheme is independent of the plaintext value, yields a great computing saving. The resemblance between Walsh-Hadamard and Haar matrices motivated us to compare WHE and HTE in Fig. \ref{fig:2}. Showing the behaviour of WHE and HTE for $l\in\{2,...,8\}$ proves that the latter has a remarkable superiority in terms of multiplication numbers compared with WHE for the large row numbers.
\subsection{Cryptographic Structure Vulnerability}
In this section, we investigate the effect of destructive factors on the structure of PHE, WHE and HTE methods.
\subsubsection{Vulnerability of PHE}
Referring to line $9$ of Algorithm \ref{algo1}, $g$, $x$, $r$ and $n$ are the key parameters in the PHE scheme. Assuming $n_1=n_2$ and $g_1=g_2$, multiplication of two ciphertexts $E(x_1)$ and $E(x_2)$ affected by an additive error (AE) is computed as follows:
\begin{multline}\label{eq:15}
E(x_1+\acute{x_1}) \cdot E(x_2+\acute{x_2}) = {g^{{x_1}+{\acute{x_1}}+{x_2}+{\acute{x_2}}}}\cdot {{({r_1}\cdot{r_2})}^{{n}+{\acute{n}}}}\\+ \Re \mod {{(n+\acute{n}})^2},
\end{multline}
where $(\acute{\cdot})$ and $\Re$ denote the AE to $(\cdot)$, and the terms consisting of AE as base numbers, respectively.

\emph{Proof 3:} $E(x_1+\acute{x_1}) \cdot E(x_2+\acute{x_2}) = {{(g+\acute{g})}^{{x_1}+{\acute{x_1}}}}\cdot {{({r_1}+\acute{r_1})}^{{n}+{\acute{n}}}}\cdot{{(g+\acute{g})}^{{x_2}+{\acute{x_2}}}}\cdot {{({r_2}+\acute{r_2})}^{{n}+{\acute{n}}}}\mod {{(n+\acute{n}})^2}={(g+\acute{g})}^{x_1}\cdot{(g+\acute{g})}^{\acute{{x_1}}}\cdot {({r_1}+{\acute{r_1}})}^{n}\cdot{({r_1}+{\acute{r_1}})}^{\acute{n}}\cdot{(g+\acute{g})}^{x_2}\cdot{(g+\acute{g})}^{\acute{{x_2}}}\cdot {({r_2}+{\acute{r_2}})}^{n}\cdot{({r_2}+{\acute{r_2}})}^{\acute{n}}\mod {{(n+\acute{n}})^2}=({g^{x_1}}+{\acute{g}^{x_1}}+\ddot{g})\cdot({g^{\acute{{x_1}}}}+{\acute{g}^{\acute{{x_1}}}}+\ddot{\tilde{g}})\cdot
({{r_1}^{n}}+{\acute{{r_1}}^{n}}+\ddot{{r_1}})\cdot({{r_1}^{\acute{{n}}}}+{\acute{{r_1}}^{\acute{{n}}}}+\ddot{\tilde{{r_1}}})\cdot({g^{x_2}}+{\acute{g}^{x_2}}+\ddot{g})\cdot({g^{\acute{{x_2}}}}+{\acute{g}^{\acute{{x_2}}}}+\ddot{\tilde{g}})\cdot
({{r_2}^{n}}+{\acute{{r_2}}^{n}}+\ddot{{r_2}})\cdot({{r_2}^{\acute{{n}}}}+{\acute{{r_2}}^{\acute{{n}}}}+\ddot{\tilde{{r_2}}})\mod {{(n+\acute{n}})^2}=({g^{{x_1}+{\acute{x_1}}+{x_2}+{\acute{x_2}}}}\cdot {{({r_1}\cdot{r_2})}^{{n}+{\acute{n}}}})+ ({{\acute{g}}^{{x_1}+{\acute{x_1}}+{x_2}+{\acute{x_2}}}}\cdot {{{(\acute{r_1}\cdot{\acute{r_2}})}}^{{n}+{\acute{n}}}}+\ldots \mod {{(n+\acute{n}})^2}\simeq E(x_1+\acute{x_1}+x_2+\acute{x_2}),$
where $\{\ddot{{\cdot}}\}$ and $\{\ddot{\tilde{{\cdot}}}\}$ are the expanded terms using binomial theorem \cite{YANG201731} where the exponent is a true data and AE, respectively.

Referring to \ref{eq:15} and $Proof$ $3$, adding error to the every possible element of PHE yields various extra computations consisting of AE as the base and/or exponent number. Furthermore, inaccurate plaintext-plaintext addition coming from disturbed ciphertext-ciphertext multiplication disrupts the process mining schemes. Therefore, in the presence of AE, increasing the number of plaintexts (i.e., $x_1$, $x_2$, \ldots), makes PHE an ultra-high computational complexity approach with inaccurate computations which is unsuccessful in the high-speed applications.
\subsubsection{Vulnerability of WHE and HTE}
Referring to \ref{eq:5} and \ref{eq:70}, the top row and the left-most column of the Walsh-Hadamard, and the top row of the Haar matrix only consist of $+1$. Furthermore, ignoring the top row and the left-most column of the former provides the cyclic shift property of rows. Since the destructive factor usually causes only a phase shifting in the aforementioned matrices \cite{KAZEMIAN}, the deterministic structure of WH and Haar matrices can be restored simply. Hence, the deterministic and low-complexity structure of the Haar matrix together with a private key is an efficient solution in the high-speed applications.


\section{Conclusion}\label{defin4}
Emergence of real-time and virtual networks into the digital business technology motivated us to investigate on the privacy-preserving concept in the industry. Since these networks are highly complex inherently, in this paper we proposed a very low-complexity encryption approach called HTE to anonymize the data before execution of the process mining techniques. Encryption with a low-complexity and easy-restorable structure in a business network including large number of organizations was the main goal of this paper. The proposed method is based on a Haar transform and one private key that is applicable to any data type in an event log. The complexity analysis proves that HTE is superior to the WHE and PHE cryptosystems in terms of overall computational complexity, significantly. The proposed method is beneficial for the companies, those privacy of consumers, complexity, cost and the speed are some of their serious concerns. The future target of the present work is to design a PM algorithm that can work with HTE encrypted data.


\ifCLASSOPTIONcaptionsoff
  \newpage
\fi
\bibliographystyle{IEEEtran}

\bibliography{keylatex}

\end{document}